\documentclass[12pt,preprint]{aastex}

\newcommand{\be}{\begin{equation}}
\newcommand{\ee}{\end{equation}}
\newcommand{\ba}{\begin{eqnarray}}
\newcommand{\ea}{\end{eqnarray}}

\newcommand{\rhom}{\rho_M}

\newcommand{\rhocard}{\rho_{\rm Card}}
\newcommand{\zcard}{z_{\rm Card}}

\newcommand{\nature}[3]{Nature {\bf B#1}, #3 (#2)}

\shorttitle{Future Type Ia Supernova Data as Tests of Dark Energy}
\shortauthors{Wang, Freese, Gondolo, Lewis}

\begin{document}

\title{Future Type Ia Supernova Data as Tests of\\
Dark Energy from Modified Friedmann Equations}
\author{Yun Wang{\footnote{Department of Physics \& Astronomy,
University of Oklahoma, Norman, OK 73019}},
Katherine Freese{\footnote{Michigan Center for Theoretical Physics,
Physics Department,
University of Michigan,
Ann Arbor, MI 48109, USA, and
Kavli Institute for Theoretical Physics, University of California,
Santa Barbara, CA 93106}}, 
Paolo Gondolo{\footnote{Department of Physics, 
Case Western Reserve University, 10900 Euclid Ave, Cleveland, OH 44106-7079}}, 
and Matthew Lewis{\footnote{Michigan Center for Theoretical Physics,
Physics Department,
University of Michigan,
Ann Arbor, MI 48109, USA}}
}

\begin{abstract}
In the Cardassian model, dark energy density arises from
modifications to the Friedmann equation, which
becomes $H^2 = g(\rhom)$, where
$g(\rhom)$ is a new function of the energy density.
The universe is flat, matter dominated, and accelerating.
The distance redshift relation predictions of generalized
Cardassian models can be very different from generic quintessence
models, and can be differentiated with data from upcoming pencil beam 
surveys of Type Ia Supernovae such as SNAP.  We have found the 
interesting result that, once $\Omega_m$ is
known to 10\% accuracy, SNAP will be able to determine the sign of the
time dependence of the dark energy density.  Knowledge of this sign
(which is related to the weak energy condition) will provide a first
discrimination between various cosmological models that fit the
current observational data (cosmological constant, quintessence,
Cardassian expansion).  Further, we have performed Monte Carlo
simulations to illustrate how well one can reproduce the form of the
dark energy density with SNAP.

To be concrete we study a class of two parameter
($n$,$q$) generalized Cardassian models that includes the original
Cardassian model (parametrized by $n$ only) as a special case.  
Examples are given of MP Cardassian models that
fit current supernovae and CMB data, and prospects for differentiating
between MP Cardassian and other models in future data are discussed.
We also note that some Cardassian
models can satisfy the weak energy condition $w>-1$ even with a dark
energy component that has an effective equation of state $w_X < -1$.

\end{abstract}



\section{Introduction}

Recent observations of Type Ia Supernovae \citep{SN1,SN2} as well as
concordance with other observations (including the microwave background
and galaxy power spectra) indicate that the universe is accelerating.
Many authors have explored a cosmological constant, a decaying vacuum
energy \citep{fafm,peebles88,frieman}, quintessence
\citep{stein,caldwell98,huey}, and gravitational
leakage into extra dimensions \citep{ddg} as possible explanations for such
an acceleration.
Recently \cite{freeselewis} proposed Cardassian
expansion as an explanation for
acceleration which invokes no vacuum energy whatsoever.\footnote{The name
Cardassian refers to a humanoid race in Star Trek whose goal is to
accelerate expansion of their evil empire.  This race looks alien to
us and yet is made entirely of matter.}   In this model
the universe is flat and accelerating, and yet consists only of matter
and radiation. 

In Cardassian models, the Friedmann equation is modified from $H^2 = 8\pi
\rho/(3m_{pl}^2)$ to 
\begin{equation}
\label{eq:mf}
H^2 = g(\rhom),
\end{equation}
where $g(\rhom)$ is a different function of the energy density,
$\rhom$ contains only matter and radiation (no vacuum), $H=\dot
a/a$ is the Hubble constant (as a function of time), and $a$ is the
scale factor of the universe. Models with gravitational
leakage into extra dimensions also give a modified Friedmann equation that can
be cast into the form of Eq.~(\ref{eq:mf}), but with a vacuum component.

 The function $g(\rhom)$ returns to the
usual $8\pi \rhom/(3m_{pl}^2)$ during the early history of the
universe, but takes a different form that drives an accelerated
expansion after a redshift $z \sim 1$.  Such modifications to the
Friedmann equation may arise, e.g., as a consequence of our
observable universe living as a 3-dimensional brane in a higher
dimensional universe \citep{cf}.  Alternatively, such a Friedmann
equation may arise if there is dark matter with self-interactions
characterized by negative pressure \citep{gondolo}.

We wish to study the detectability of the altered Friedmann equations
by upcoming observations of Type Ia Supernova such as SNAP.  As discussed in
\cite{freese}, the redshift distance relationship predictions
for generalized Cardassian models can be quite different from
generalized quintessence models.  It is the goal of this 
paper to see how well we can reproduce the correct form of
this dark energy density in upcoming experiments.

For concreteness, we investigate a particular version of generalized
Cardassian cosmology. However, our results are intended to 
generalize to any Cardassian cosmology, i.e., to any function
$g(\rhom)$.  The particular model we study is meant to illustrate
that, generically, modified Friedmann equations can lead to
specific detectable predictions in experiments like SNAP.

In this paper we study the following generalization of the original 
Cardassian model:
\begin{equation}
\label{eq:FRW}
H^2 = \frac{8\pi G \rhom}{3} \, 
\left[1+ \left( \frac{\rhocard}{\rhom}\right)^{q(1-n)}
\right]^{1/q} .
\end{equation} We call this model ``Modified Polytropic Cardassian'' (MP
Cardassian)\footnote{The name ``Modified Polytropic'' arises in the
context of treating this model as a fluid; then the relationship
between energy density and pressure is roughly polytropic (see
\cite{gondolo}).}  \citep{gondolo},
where $G=1/m_{pl}^2$ is Newton's universal 
gravitation constant, $\rho_{\rm Card}$ is a characteristic constant 
energy density, and where we
take $n<2/3$ and $q>0$.
The original power law Cardassian model corresponds to $q=1$.

For comparison,
we remind the reader of the original power law 
Cardassian model that was proposed in 
\cite{freeselewis}, which had the following specific form of $g(\rhom)$:
\begin{equation}
\label{eq:new}
H^2 = {8\pi\over 3 m_{pl}^2} \rhom + B \rhom^n \,\,\,\,\,\, {\rm with}
\,\,\,\,\,\, n<2/3 .
\end{equation}
This is equivalent to writing
\begin{equation}
  \label{eq:friedcard}
  H^2 = \frac{ 8 \pi G}{3} \, \rhom \left[ 1 + 
    \left( \frac{\rhocard}{\rhom} \right)^{\!\!1-n\,} \right] ,
\end{equation}

The first term inside the
bracket of Eq.(\ref{eq:FRW}) and Eq.(\ref{eq:friedcard})
dominates initially, so that ordinary Friedmann Robertson Walker (FRW)
behavior takes place throughout the early universe.  At a redshift 
$\zcard \sim 1$,
the two terms inside the bracket become equal, and henceforth the second term
dominates.  Once the second term dominates, it drives the universe
to accelerate.
The energy density at which the two terms become equal is
$\rhocard=\rho_0 (1+\zcard)^3$, where $\rho_0$ is the matter density today. 
The MP Cardassian model of Eq.(\ref{eq:FRW})
depends on three parameters: the numbers $n$ and $q$ and
the density $\rhocard$. 
The latter can be traded for the observed matter mass
density $\Omega_m^{obs}$ (see Eq.~[\ref{eq:rhocard}] below).

The original power law 
Cardassian model gave the same distance-redshift relation
as a quintessence model with constant equation of state parameter $w_q
=n-1$.  Generalized Cardassian models, on the other hand, give
predictions for the distance-redshift relation that can be very
different from generic quintessence models. For example, some Cardassian
models can satisfy the weak energy condition $w>-1$ even with a dark
energy component that has an effective equation of state $w_X = p_X/\rho_X
< -1$.  Note that an effective
$w_X <-1$ is consistent with recent CMB and large scale structure
data \citep{Schue02,Mel02}.
In this paper we explore these differences and their testability for
the MP Cardassian model.

The Cardassian model also has the attractive feature that matter alone
is sufficient to provide a flat geometry. Because of the extra term on the
right hand side of the Friedmann equation, the 
critical mass density necessary to have a flat universe 
can be modified, e.g.\ to 0.3 of the usual value.
Hence the matter mass density can have exactly this new critical value and
satisfy all the observational constraints such as given by the baryon
cluster fraction and the galaxy power spectrum.

{\it General Notation:}

In a flat universe,
the total energy density of the universe can be written as
\be
\label{eq:rho_tot}
\rho_{total}(z)= \rho_{c,\rm old} \left[ \Omega_m^{obs} (1+z)^3 +
\Omega_X f_X(z)\right] .  \ee Here, $\Omega_m^{obs}$ is the observed
matter density of the universe; we will take $\Omega_m^{obs} = 0.3$ as
our fiducial value.  The critical density of the universe (for the
ordinary Friedmann equation) is $\rho_{c,old} = 2 \times 10^{-29}
h_0^2$ gm/cm$^3$, where $h_0$ is the Hubble constant today in units of
100 km/s/Mpc.  We take 
\begin{equation}
\Omega_X=1-\Omega_m^{obs}
\end{equation}
 (for a flat
universe with total energy density $\Omega_{tot}=1$), and
$f_X(z=0)=1$.  The subscript $X$ refers to any component of the
universe that provides an additional term in Einstein's equation;
generically it is called ``dark energy'', but in the Cardassian case
it is an additional matter term.  The dark energy density is
\begin{equation}
\label{eq:f}
\rho_X(z)=\rho_X(0) f_X(z) = \rho_{c,\rm old} \Omega_X f_X(z) .
\end{equation}
If the dark energy density corresponds to a cosmological
constant, then one finds that
$f_X(z)=1$ at all redshifts $z$.

Note that for generalized Cardassian models,
{\it both} terms in Eq.~(\ref{eq:rho_tot}) come from matter;
\begin{equation}
\Omega^{tot}_m=\Omega_m^{obs}+ \Omega_X=1 .
\end{equation}
For the MP
generalized Cardassian models of Eq.~(\ref{eq:FRW}),
the dimensionless dark energy density $f_X(z)$ is given by
\be
f_X(z) = \frac{\rho_X(z)}{\rho_X(0)} = 
\frac{\Omega_m^{obs} (1+z)^3}{1-\Omega_m^{obs} }
\left\{ \left[ 1+ \frac{ (\Omega_m^{obs})^{-q} -1}{(1+z)^{3(1-n)q}}\right]
^{1/q} -1 \right \}.
\ee
For fixed $\Omega_m^{obs}$, it depends on the two dimensionless parameters $n$
and $q$. 

In Cardassian models,
the observed matter mass density fraction today is given by the ratio of
the critical mass density of the Cardassian universe, $\rho_{c,\rm Card} = 
\rho_0$,
and that of the standard universe, $\rho_{c,\rm old} \equiv 3H_0^2/(8 \pi
G)$. So the observed matter mass fraction today in the MP Cardassian model
is \citep{freeselewis} 
\be
\Omega_m^{obs} = \frac{\rho_0}{\rho_{c,\rm old}} = \frac{1}
{ \left[ 1+(1+\zcard)^{3q(1-n)} \right]^{1/q} } \, .
\ee 
Inversely, we can express $\zcard$, and $\rhocard$, in terms of 
$\Omega_m^{obs}$ as
\be
1+\zcard = \left[ \left( 
\frac{1}{\Omega_m^{obs}} \right)^q - 1 
\right]^{\frac{1}{3q(1-n)}} , 
\ee
and
\be
\label{eq:rhocard}
\rhocard = \rho_{c,\rm old} \Omega_m^{obs} \left[ \left( 
\frac{1}{\Omega_m^{obs}} \right)^q - 1 
\right]^{\frac{1}{q(1-n)}} .
\ee

{\it Outline:}
First, we will compare the Modified Polytropic Cardassian model
of Eq.(\ref{eq:FRW}) with existing data from supernovae and cosmic
microwave background data.

Next, we explore how plausible future SNe Ia data
can be optimally used to constrain dark energy models, and whether
generalized Cardassian models can be differentiated from
generic models of quintessence and models with a cosmological constant.

Most of this work was performed in September 2002 when all the authors
were at the Kavli Institute for Theoretical Physics in Santa Barbara.
Subsequent dispersal of all the authors to different parts of the
country caused the conclusion of the paper to take a long time.

\section{Comparison of MP Cardassian with Current Data}

In this section, we compare the Modified Polytropic Cardassian Model
of Eq.~(\ref{eq:FRW}) with current supernovae and cosmic microwave
background (CMB) data.  We will see that the existing data can be well
fit for several choices of the parameters
$n$ and $q$. 

In a smooth Friedmann-Robertson-Walker (FRW) universe,
the metric is given by $ds^2=dt^2-a^2(t)[dr^2/(1-kr^2)+r^2 (d\theta^2
+\sin^2\theta \,d\phi^2)]$, where $a(t)$ is the cosmic scale factor,
and $k$ is the global curvature parameter.
The comoving distance $r$ is given by \citep{Weinberg72}
\be
\label{eq:r(z)}
r(z)=cH_0^{-1}\, \frac{S(\kappa \Gamma)}{\kappa}, \hskip 1cm
\kappa \equiv \left| \Omega_k \right|^{1/2},
\ee
\be
\Gamma(z;\Omega_m^{obs},\Omega_X, F)=\int_0^zdz' \frac{1}{E(z')},
\ee
\be
E(z') \equiv \left[ \Omega_m^{obs}(1+z')^3+ \Omega_X\, f_X(z')+\Omega_k(1+z')^2
\right]^{1/2},
\ee
where $\Omega_k = 1-\Omega_m^{obs}- \Omega_X$, and
\ba
S(x)&=&\sinh(x), \hskip 2cm \Omega_k>0 \nonumber \\
&=&x, \hskip 3cm \Omega_k=0 \nonumber \\
&=&\sin(x), \hskip 2cm \Omega_k<0.
\ea
The angular diameter distance is given by $d_A(z)=r(z)/(1+z)$,
and the luminosity distance is given by $d_L(z)=(1+z) r(z)$.

The distance modulus for a standard candle at redshift $z$ is
\be
\label{eq:mu0p}
\mu_p (z)\equiv m-M= 5\,\log\left( \frac{ d_L(z)}{\mbox{Mpc}} \right)+25,
\ee
where $m$ and $M$ are the apparent and absolute magnitudes of the standard
candle, and $d_L(z)$ is its luminosity distance.

Type Ia supernovae (SNe Ia) are our best candidates for cosmological 
standard candles, because they can be calibrated
to have small scatters in their peak luminosity \citep{Phillips93,Riess95}.

Fig.~1 shows the measured distance modulus (actually the deviation of the
distant modulus with respect to the expected values for an open universe with
$\Omega_m^{obs}=0.3$ and $\Omega_{\Lambda}=0$) for flux-averaged\footnote{ Here
  we briefly describe flux-averaging.  Due to the inhomogeneous distribution of
  matter in our universe, the light from perfect standard candles (which all
  have exactly the same peak luminosity) at a given redshift $z$ will
  experience different amounts of bending (due to matter inhomogeneity along
  different lines of sight) before reaching the observer.  Hence even perfect
  standard candles will be observed to have a non-Gaussian spread in peak
  luminosity due to gravitational lensing
  \citep{frieman97,Wamb97,HolzWald98,ms99,Wang99,Wang02,Munshi03}.
  If this effect is not properly taken
  into account, the estimated cosmological parameters will be biased, i.e., the
  estimated mean of the parameters will deviate from the true value of the
  parameters.  Fortunately, the total number of photons from all the standard
  candles at redshift $z$ should remain unchanged in the presence of
  gravitational lensing (which only redistributes the photons by bending the
  light from each standard candle); therefore the average peak luminosity of
  all the standard candles at $z$ should be the same as the peak luminosity of
  a standard candle at $z$ {\it without} gravitational lensing.  This is the
  basic idea behind flux-averaging of type Ia supernova data. For details, see
  \cite{Wang00b}.}\citep{Wang00b} SNe Ia data \citep{SN1,SN2} as a function of
redshift.  For comparison, superposed on the data points are the predictions
for several familiar cosmological models (dotted curves, from top to bottom):
($\Omega_m^{obs}$, $\Omega_{\Lambda})=$ (0.3, 0.7); (0.3, 0); and (1, 0).  In
addition, three examples of modified polytropic Cardassian models from
Eq.~(\ref{eq:FRW}) are shown, all with $\Omega_m^{obs} = 0.3$. The three models
have parameters $n=0.2$, $q=1$ (solid curve); $n=0.2$, $q=2$ (short-dashed
curve); and $n=0.2$, $q=3$ (long-dashed curve).  Note that the solid curve is
equivalent to a quintessence model with $w_q = -0.8$.  
Also shown in Fig.1 is a quintessence model with $w_q(z)=-1+0.5z$
for comparison (dot-dashed line).
All three generalized
Cardassian models shown satisfy current constraints from SNe Ia.  So we
conclude that MP Cardassian models fit the existing supernovae data very well.

Fig.~2 shows the $(n,q)$ parameter space with constraints from current
observational data from both supernovae and the cosmic microwave
background (CMB), at a fixed value $\Omega_m=0.3$.  The MP Cardassian
model with parameters $(n,q)$ is compared with a fiducial $\Lambda$CDM
(cold dark matter) model $\Omega_m=0.3$, $\Omega_b=0.05$,
$\Omega_\Lambda=0.7$, and $h=0.65$.  The constraints are derived by
requiring that the MP Cardassian models agree with the fiducial
$\Lambda$CDM model to within 1$\sigma$ of the measurement
uncertainties of the Wilkinson Microwave Anisotropy Probe (WMAP) CMB
data \citep{Bennett03} and the current SN Ia data [\citep{SN1,SN2},
flux-averaged with $\Delta z=0.05$ \citep{Wang00b}].

The region between the thick solid lines in Fig.~2 corresponds to
$(n,q)$ values of the MP Cardassian models that satisfy the current SN
Ia data [flux-averaged with $\Delta z=0.05$] within one sigma of the
fiducial $\Lambda$CDM model, i.e., $\Delta \chi^2 = \chi^2_{MPC}-
\chi^2_{ {\Lambda}{\rm CDM} }=1$.  Note that the fiducial $\Lambda$CDM
model ($\Omega_m=0.3$, $\Omega_\Lambda=0.7$, $h=0.65$) corresponds to
$n=0$, $q=1$. Complementary plots showing constraints from the power
law Cardassian model with fixed $q=1$ and varying $n$ and $\Omega_m$
are given in \cite{SenSen,Zhu02,Zhu03}.

The dashed line in Fig.~2 indicates the constraints from 
WMAP CMB data. When the cosmological parameters are varied, 
the shift in the whole CMB angular spectrum is
determined by the shift parameter \citep{Bond97,Mel02,Odman02}
\be
{\cal R} = \sqrt{\Omega_m}\, H_0 \, r(z_{dec})
\ee
where $r(z_{dec})$ denotes the comoving distance to the decoupling
surface in a flat universe. The results from WMAP data require
$H_0 \, r(z_{dec})= 3.330 \,^{+0.193}_{-0.158}$ and hence give the
allowed range in the shift parameter.  The corresponding constraints
on MP Cardassian model parameters ($n,q$) are represented by the
dashed line in Fig.2.  All models in Figure 2 below the dashed line
lie within this allowed range for the shift parameter.  We conclude
that the Modified Polytropic Cardassian models of Eq.(\ref{eq:FRW})
are compatible with current supernova and CMB data.

\section{Comparison of Models Using Simulated Future Data}

In this section, we will construct simulated type Ia supernova
(SN Ia) data for three dark energy models: a Modified Polytropic Cardassian 
model, a cosmological constant model, and a quintessence model.
We then investigate if we can
recover the original theory from the simulated data.  In particular
we want to see if Cardassian cosmology can be differentiated
from generic quintessence models or a cosmological constant
by analysis of upcoming SN Ia data. We will
begin with a modified polytropic Cardassian model of Eq.(\ref{eq:FRW}),
choose specific values of the parameters $q$ and $n$, and
see how many SNe Ia we would expect.

The measured distance modulus for a SN Ia (labeled ``$l$'') is 
\be \mu_0^{(l)}=
\mu_p^{(l)}+\epsilon^{(l)} \ee where $\mu_p^{(l)}$ is the theoretical
prediction [see Eq.(\ref{eq:mu0p})], and $\epsilon^{(l)}$ is the uncertainty in
the measurement, including observational errors and intrinsic scatters in the
SN Ia absolute magnitudes.  In the simulated data set, we take the dispersion
in SN Ia peak luminosity to be $\Delta m_{int}=0.16\,$mag (this is the rms
variance of $\epsilon^{(l)}$).

Many (but not all) models of dark energy can be characterized
by an equation of state $w_X(z) = p_X(z) / \rho_X(z)$, where $p_X(z)$
is the pressure.  Most authors have concentrated on constraining
the equation of state $w_X$ of the dark energy from SN data.
However, it was shown by \cite{Maor01,barger} that it is extremely hard
to constrain $w_X$ using SN data.  Instead,  
\cite{Wang01a}
emphasized that it is easier to extract information on the dark
energy density $\rho_X(z)$, instead of $w_X(z)$, from the data.
This is because there are multiple integrals relating
$w_X(z)$ to the luminosity distance $d_L(z)$ of SN, which
results in a ``smearing'' that obscures the difference between
different $w_X(z)$.  It is better to use $\rho_X(z)$ directly,
as it is related to the time derivative of the comoving distance
to SN Ia, $r'(z)$; hence it is less affected by the smearing
effect.  The advantage of measuring $\rho_X(z)$ over measuring 
$w_X(z)$ was confirmed by \cite{tegmark}.
In their work, \cite{Wang01a} assumed that $\rho_X'(z) > 0$,
a condition equivalent to the weak energy condition for those
cases in which the ordinary Friedmann equation applies.
Here, on the other hand, we make no such assumption. 
In fact, 
we will
show that it is possible to determine the sign of $\rho_X'(z)$.

\subsection{Determining the sign of the time dependence of dark energy
density, with prior on $\Omega_m$}

We simulate data for three models (see Table 1):
(1) a cosmological constant model;
(2) a MP Cardassian model with $n=0.2$ and $q=2$,
which has $\rho_X'(z) \leq 0$;
(3) a quintessence model with $w_X(z)=-1+0.5z$,
which has $\rho_X'(z) \geq 0$.

\begin{center}
Table 1\\
{\footnotesize{Dark Energy Models}}

{\footnotesize
\begin{tabular}{|l|l|l||}
\hline 
Model & model parameters & $\rho_X'(z)$ \\  
\hline 
$\Lambda$CDM    &  $\Omega_m=0.3$, $\Omega_{\Lambda}=0.7$   & 
$\rho_X'(z)=0 $\\
MP Cardassian model  & $\Omega_m^{obs}=0.3$, $n=0.2$, $q=2$ &
$\rho_X'(z) < 0 $ \\
quintessence model  & $\Omega_m=0.3$, $w_q(z)=-1+0.5\,z$   & 
$\rho_X'(z)\geq 0 $ \\
\hline 
\end{tabular}
}
\end{center}

Note that MP Cardassian models can have either $\rho_X'(z) \geq 0$ or
$\rho_X'(z) < 0$.  On the other hand, popular quintessence models in which the
quintessence field tracks the matter field have $\rho_X'(z) \geq 0$
\citep{barger}.  Fig.~3 shows the sign of $\rho_X'(z)$ in the $(n,q)$ parameter
space at fixed $\Omega_m=0.3$ for MP Cardassian models [see Eq.(\ref{eq:FRW})]
with $\Omega_m^{obs}=0.3$.  The arrows indicate the regions in which
$\rho_X'(z) \geq 0$ and $\rho_X'(z) < 0$ respectively.  The region in between
indicates models with dark energy densities that are {\it not} monotonic
functions of time.

Note that the weak energy condition requires that the total equation 
of state $w \geq -1$. In the context of treating Cardassian models as a fluid
\citep{gondolo}, this is 
\be
w = \frac{p_X}{\rho_m +\rho_X} = 
- \frac{ \left[ (\Omega_m^{obs})^{-q} - 1 \right]\, (1-n)}
{ (1+z)^{3(1-n)q} + (\Omega_m^{obs})^{-q} - 1 },
\ee
which satisfies $w \geq -1$ for 
the parameter choices we are interested in:
$\Omega_m^{obs} <1$, $n<1$, and $q>1$.
Therefore, all viable MP Cardassian models satisfy the weak energy
condition but can have $w_X < -1$. Note an effective
$w_X <-1$ is consistent with recent CMB and large scale structure
data \citep{Schue02,Mel02}.

Some scalar field dark energy models with $w_X < -1$ 
have been studied previously
\citep{caldwell}; models which are stable
despite violating variants of the weak energy conditions
 have been found to be difficult to construct
\citep{Carroll03}.  Our proposal here is a different alternative
to the models previously studied.

We will now assume that the data set is given, either
from the simulated data sets described above, or, in the future
from SNAP\footnote{Note that the current SNAP design
is sustantially improved than before \citep{Tarle02}. 
Here we assume that SNAP will obtain all SNe Ia 
in its survey fields up to
$z=1.7$, similar to a supernova pencil beam survey \citep{Wang00a,Wang01b}.}.
  We will show that investigation of the
data set can reproduce the sign of the time-dependence
of the dark energy density, assuming one knows the matter density
to an accuracy of 10\%.

Given our data set, we now proceed as though we have
no information on where it comes from; i.e., we
proceed as though we did not know which model it came from.
We parametrize the dark energy density in
order to allow us to compare it to the data set.
We take $\rho_X(z)$
to be an arbitrary function.  To approximate the function,
we parametrize it by its value
at $n_{bin}$ equally spaced redshift values, $z_i$,
$i=1$,2,...,$n_{bin}$, $z_{n_{bin}}=z_{max}$.
The value of $\rho_X(z)$ at other redshifts are given by
linear interpolation, i.e., 
\ba
\label{eq:f(z)}
&&\rho_X(z)=\left( \frac{z_i-z}{z_i-z_{i-1}} \right)\, \rho_{i-1}+
\left( \frac{z-z_{i-1}}{z_i-z_{i-1}} \right)\, \rho_i,
 \hskip 1cm z_{i-1} < z \leq z_i, \nonumber \\
&& z_0=0, \,\, z_{n_{bin}}=z_{max} .
\ea
The values of the dark energy density
$\rho_i$ ($i=1,2,...,n_{bin})$ are the independent variables
to be estimated from data; note that the number of independent
variables is $n_{bin}$.    Again, we proceed as though we
had absolutely no information on the function $\rho_X(z)$,
and treat it as a completely arbitrary function.

The complete set of parameters, then, is
\begin{equation}
{\bf s} \equiv (\Omega_m^{obs},\,\,\, \rho_i, \,\,\,{\rm and} \,\,\, n_{bin}) ,
\end{equation}
where $i = 1,...,n_{bin}$ as described above.
Hence our number of parameters is $N=n_{bin}+2$.
We will vary the number of bins $n_{bin}$ between 1 and 10,
and look for the optimal fit to the data. To illustrate,
an arbitrary function may become a good approximation to
the data for 4 bins whereas it is a miserable fit for 3 bins.

We expand the adaptive iteration method developed
in \cite{Wang01a} and \cite{Wang01b};
unlike what is done in those papers, we do not restrict
ourselves to cases where $\rho_X'(z) > 0$.

We can now determine a best fit to the set of parameters $\bf{s}$
by using a $\chi^2$ statistic, with
\citep{SN1}
\be
\chi^2(\mbox{\bf s})=
\sum_l \frac{ \left[ \mu^{(l)}_p(z_l| \mbox{\bf s})-
\mu_0^{(l)}(z_l) \right]^2 }{\sigma_l^2 },
\ee
where $\mu^{(l)}_p(z_l| \mbox{\bf s})$ is the prediction
for the distance modulus at redshift $z_l$, given the set
of parameters $\bf{s}$.  Here
$\sigma_l$ is the dispersion of the measured distance modulus
due to intrinsic and observational
uncertainties in SN Ia peak luminosity.

To reduce the computation time, we can integrate over
the Hubble constant $H_0$ analytically, and define a modified
$\chi^2$ statistic, with
\be
\label{eq:chi2mod}
\tilde{\chi}^2 \equiv \chi_*^2 - \frac{C_1}{C_2} \left( C_1+ 
\frac{2}{5}\,\ln 10 \right) - 2 \ln\,h^*,
\ee
where $h^*$ is a fiducial value of the dimensionless Hubble constant $h$,
\be
\mu_p^* \equiv \mu_p(h=h^*)=42.384-5\log h^*+ 5\log \left[H_0 r(1+z)\right],
\ee
and
\ba
\chi_*^2 &\equiv& \sum_l \frac{1}{\sigma_l^2} \left( \mu_{p}^{*(l)}-
\mu_{0}^{(l)} \right)^2, \nonumber \\
C_1 &\equiv& \sum_l \frac{1}{\sigma_l^2} \left( \mu_{p}^{*(l)}-
\mu_{0}^{(l)} \right), \nonumber \\
C_2 &\equiv& \sum_l \frac{1}{\sigma_l^2} .
\ea
It is straightforward to check that the derivative of $\tilde{\chi}^2$
with respect to $h^*$ is zero; hence our results are independent of the 
choice of $h^*$. We take $h^*=0.65$. 

For a given choice of $n_{bin}$, we can minimize the
modified $\chi^2$ statistic of Eq.(\ref{eq:chi2mod}) to find the best
fit $\Omega_m^{obs}$ and $\rho_X(z)$ (parametrized by $\rho_i$, $i=1$, 2, ...,
$n_{bin}$).  We can find one sigma error bars by finding values
with $(\Delta \chi)^2 = 1$ from the minimum.

For each model in Table 1, we obtain {\it four} sets of best fit parameters.
We apply four different constraints to the arbitrary function $\rho_X(z)$
in order to discover which one allows a good fit.
The four constraints are: \hfill\break
(i) $\rho_X(z)=\rho_X(0)
= {\rm constant}$; i.e.,
a cosmological constant model; \hfill\break
(ii) $\rho_X'(z) \geq 0$; \hfill\break
(iii) $\rho_X'(z) < 0$; \hfill\break
and (iv) completely unconstrained $\rho_X(z)$. \hfill\break
For each of these constraints, we find the best fit parameters.

Figure 4 shows our results: panels (a) and (b)
correspond to the MP Cardassian and quintessence models described in Table 1
respectively.
For simulated data of Model 1 of Table 1, a cosmological constant model, 
we find that
$\Omega_m^{obs}$ is estimated correctly
to 1\%
accuracy, regardless of the assumption made about $\rho_X'(z)$.
For Model 2 (MP Cardassian) and Model 3 (quintessence), Fig.4(a) and (b)
show the best fit $\Omega_m^{obs}$,
under all of the four constraints above,
for $n_{bin}$ values ranging from 1 to 10. 
We find that assuming the wrong sign for $\rho'_X(z)$ leads to an estimated
$\Omega_m$ that differs from the assumed $\Omega_m^{obs}$ by more than 10\%.
 The different constraints on the sign of $\rho_X'(z)$
are represented by different point types.
The solid horizontal line
is our fiducial value of $\Omega_m = 0.3$ (i.e., we are assuming
that that this is the true 
value of the matter density), and the dot-dashed horizontal
lines indicate 10\% error bars about his fiducial value. We are assuming
that $\Omega_m$ is known to within 10\% from other data sets.

These plots are {\it not} intended to emphasize the dependence of
$\Omega_m^{obs}$ on $n_{bin}$.  Indeed, as discussed above, the reason that we
have found the best fit $\Omega_m$ for a variety of $n_{bin}$ values is simply
that the parametrization of the arbitrary function $\rho_X(z)$ may be poor for
one value of $n_{bin}$ but excellent for another; we take a given model to be a
good one if it lies within the 10\% range on $\Omega_m$ for several values of
$n_{bin}$.

Fig.4(a) and (b) show estimated $\Omega_m^{obs}$ as function of $n_{bin}$ 
for Model 2 (MP
Cardassian model) and Model 3 (a quintessence model).
In Fig.4(a), only the $\Omega_m^{obs}$ values estimated
assuming that $\rho_X'(z) \leq 0$ consistently (i.e., for 
most values of $n_{bin}$) lie within
10\% of the true value of $\Omega_m^{obs}=0.3$.
Hence, we have indeed recovered
the correct general time-dependence
of the model underlying this set of simulated data.
In Fig.4(b), only the $\Omega_m^{obs}$ values estimated
assuming that $\rho_X'(z) \geq 0$ 
deviate by less than 10\% 
from the true value of $\Omega_m^{obs}=0.3$.
Again, we have recovered the correct general time-dependence
of the model underlying this set of simulated data.
For all three models, we have been able to correctly 
ascertain the sign of $\rho_X'(z)$ with this technique.

This indicates that the estimated $\Omega_m^{obs}$ 
(for a variety of values of $n_{bin}$),
together with a 10\% accurate prior on $\Omega_m^{obs}$,
can be used to determine the general time-dependence
of the dimensionless dark energy density; i.e.,
the sign of $\rho_X'(z)$.


\subsection{Estimating dark energy density from data}

In this section we create a large number of Monte Carlo samples
to see how well we can reconstruct the entire function $\rho_X(z)$.
To create these samples, we need first to identify the best-fit model 
as follows.

%

For a given data set, we choose the best-fit model
with $n_{bin}$ [the number of parameters used to parametrize 
the dimensionless dark energy density $\rho_X(z)$] that satisfies
three conditions:\\
(1) it corresponds to an estimated $\Omega_m$ value that deviates 
    less than 10\% from the true value;\\
(2) as we decrease $n_{bin}$ from a large value, say, $n_{bin}=10$, 
        it minimizes the $\chi^2$ per degree 
        of freedom, $\chi^2_{pdf}=\chi^2/(N_{data}-\nu)$, without 
        significantly shifting the estimated value of 
        $\Omega_m$ \citep{Wang01b}.
        $N_{data}$ is the number of SNe Ia, and $\nu$ is the number of 
        parameters estimated from data.
        Long dashed lines in Fig.4 show $\chi^2_{pdf}$
        as function of $n_{bin}$ on an arbitrary scale\footnote{The scale is 
        adjusted for the curve to fit in the figure.};\\
(3) if $\rho_X'(z) \neq 0$, then $n_{bin}>1$.

Now we apply the above conditions to Fig.4(a) and (b).
In Fig.4(a), as we decrease $n_{bin}$ from $n_{bin}=10$,
the significant shifting in the estimated $\Omega_m$ occurs at
$n_{bin}=6$, which also has the smallest $\chi^2_{pdf}$ 
for $6\leq n_{bin} \leq 10$.
We find that for the MP Cardassian model, 
SNAP data yield an optimal $n_{bin}=6$.
In Fig.4(b), as we decrease $n_{bin}$ from $n_{bin}=10$,
the significant shifting in the estimated $\Omega_m$ occurs at
$n_{bin}=3$, which also has the smallest $\chi^2_{pdf}$ 
for $3\leq n_{bin} \leq 10$.
Hence for the quintessence model,
SNAP data yield an optimal $n_{bin}=3$.

To derive the error distribution of estimated parameters
$\Omega_m^{obs}$ and $\rho_i$ ($i=1$, 2, ..., $n_{bin}$, see
Eq.(\ref{eq:f(z)})), we create $10^4$ Monte Carlo samples 
by adding dispersion in peak luminosity of $\Delta m_{int}=0.16$ mag to
the distance modulus $\mu_p(z)$ [see Eq.(\ref{eq:mu0p})]
predicted by the best-fit model (i.e., assuming that
the best-fit model is the true model). 
This is equivalent to making $10^4$ new ``observations'',
each similar to the original data set \citep{Press}.
The same analysis used
to obtain the best-fit model from the data is performed
on each Monte Carlo sample. We use
the distribution of the resultant
estimates of the parameters ($\Omega_m^{obs}$ and $\rho_i$)
to derive the mean and 68.3\% and 99.73\% confidence level intervals
of the estimated parameters.
Wang \& Lovelace (2001) showed that such a Monte Carlo analysis gives 
less biased estimates of parameters than a maximum likelihood analysis,
i.e., the Monte Carlo mean of 
estimated parameters deviate less from the true values of the parameters.

Fig.~5 shows the estimated dimensionless dark energy density
$\rho_X(z)$ for a generalized Cardassian model with $n=0.2$ and $q=2$
from simulated SN data from SNAP assuming that we know 
$\Omega_m$ to ~10\% accuracy.
The solid line indicates the underlying true model for $\rho_X(z)$.
The horizontal dashed line near the top of the figure
indicates $\rho_X(z)=\rho_X(0) = {\rm constant}$, 
a cosmological constant model.
The horizontal dotted line near the bottom of the figure
indicates $\rho_X(z)=0$. We impose $\rho_X(z) \geq 0$.
The estimated $\Omega_m$ values (obtained by reconstructing the
model from the Monte Carlo samples)
are listed at the bottom of the plot with 68.3\% confidence level 
intervals.  
Where the actual value of $\Omega_m$ for the fake data
set was $\Omega_m = 0.3$, we see that the reproduced $\Omega_m$
from our Monte Carlo study is $\Omega_m = 0.298 (-0.023, + 0.024)$.
Indeed this reproduced value lies within 10\% of the correct
$\Omega_m$.  

The error bars of the reproduced estimates of 
$\rho_X(z)$ have been computed using
$10^4$ Monte Carlo random samples derived from the simulated data.
The solid error bars and the dotted error bars 
indicate the 68.3\% and 99.73\% confidence level intervals respectively.
Hence, from Fig.~5, we see that a MP Cardassian model with a set of
parameters that fit the current observational data,
$\Omega_m^{obs}=0.3$, $n=0.2$, $q=2$ ,
can be differentiated from a cosmological constant model at 
99.73\% confidence level.
We have shown the accuracy with which one can reconstruct the
form of the dark energy density.  We see that SNAP can indeed
differentiate between different models.

\section{Discussion and Conclusion}

We have compared a particular form of Cardassian model, the Modified Polytropic
Cardassian model of Eq.~(\ref{eq:FRW}), with existing data from supernovae and
cosmic microwave background measurements. We have found that current data
constrain the parameter space of the MP Cardassian model.

We have shown that future type Ia
supernova (SN Ia) data from SNAP can differentiate various dark energy models 
(cosmological constant, quintessence, and generalized Cardassian 
expansion), assuming that $\Omega_m$ is known to 10\% accuracy.
We have found the interesting result 
that the sign of the time dependence of the dark
energy density can be determined by SNAP.

Further, we have performed Monte Carlo samples to illustrate
how well one can reproduce the form of the dark energy density with SNAP.
For example, a MP Cardassian model with a set of
parameters that fit the current observational data,
$\Omega_m^{obs}=0.3$, $n=0.2$, $q=2$ ,
can be differentiated from a cosmological constant model at 
99.73\% confidence level.

We wish to remark on another test of generalized Cardassian models.
There are two independent motivations for these models: 1) they may
arise as a consequence of imbedding our observable universe as a
3-brane in higher dimensions (see, e.g., \cite{cf}), and 2) these
models may be described in terms of a fluid interpretation, in which
the dark energy density may be due to self-interaction of dark matter
particles \citep{gondolo}. In this second interpretation, we have
developed a fully relativistic treatment of the resultant modified
Euler's equations, Poisson equations, and energy conservation.  Then
we analyzed \citep{gondolo} the linear growth of density fluctuations
in the fluid interpretation.  A more complete study of perturbation
growth is in progress.  Of particular interest is the study of the
Integrated Sachs Wolfe effect in the Cosmic Microwave Background.  It
is possible that the deficit of power on large angular scales (low
order multipoles) may be explained in generalized Cardassian models.

\acknowledgements
We thank the Kavli Institute for Theoretical Physics at
the University of California, Santa Barbara, for hospitality.  
Most of this work was performed in September 2002
when all the authors were at the Institute for Theoretical Physics in
Santa Barbara.  Subsequent dispersal of all the authors to different
parts of the country caused a delay in the publication of this paper.
 
It is a pleasure for us to thank Arlin Crotts, Greg Tarle, the referee, 
and especially Josh Frieman for helpful
comments.  This research was supported in part by the National Science
Foundation under Grant No.\ PHY99-07949, 
NSF CAREER grant AST-0094335 (Y.W.),
the Department of Energy grant at the University of Michigan
 (K.F.  and M.L.),
the Michigan Center for Theoretical Physics (K.F. and M.L.). 
K.F. thanks the
Aspen Center for Physics, where part of this research was conducted, for
hospitality during her stay.

\clearpage
\setcounter{figure}{0}

\figcaption[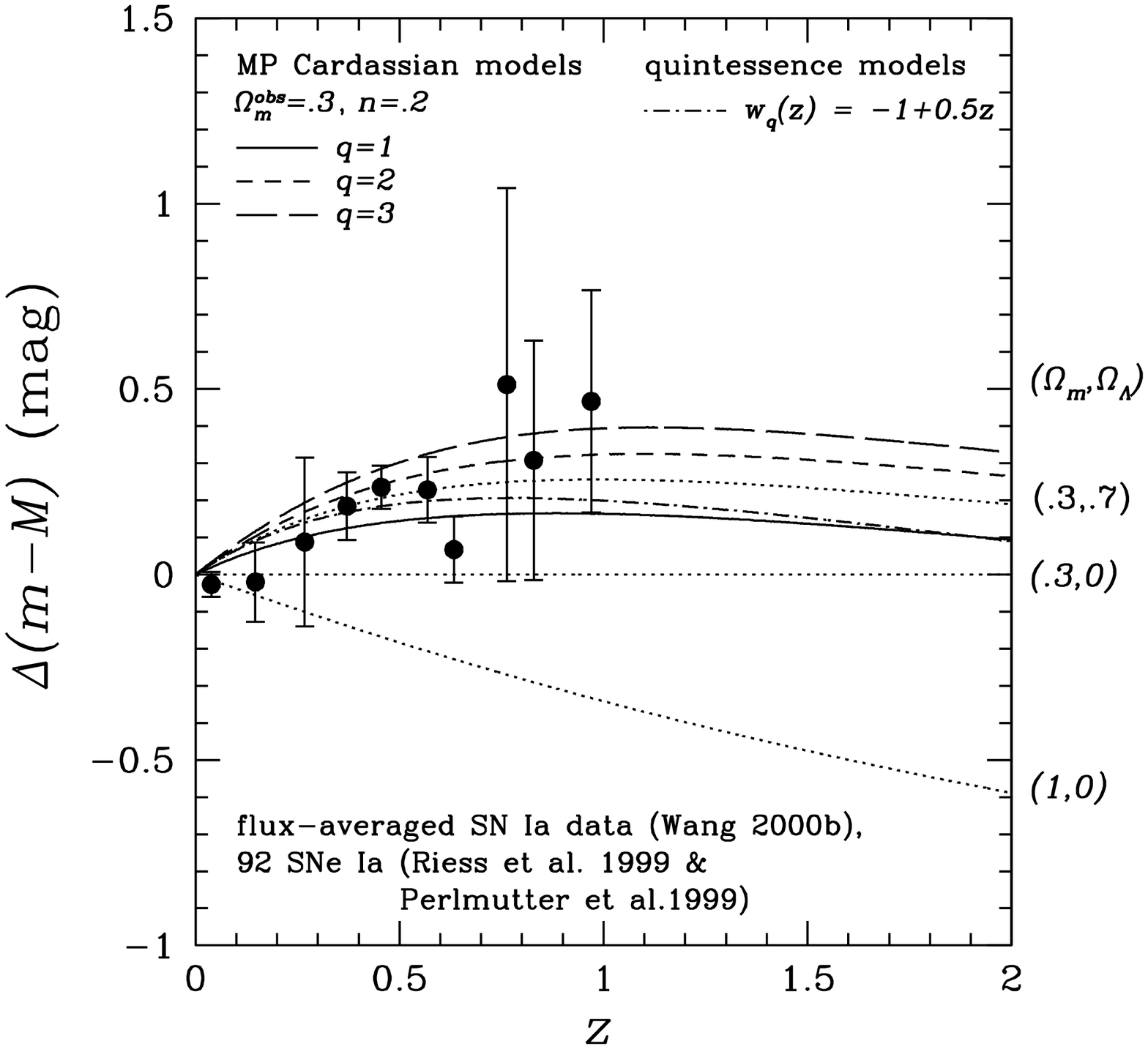]
{Examples of MP Cardassian models [see
Eq.(\ref{eq:FRW})] that satisfy current observational
constraints from type Ia
supernovae (SNe Ia) data.
Three MP Cardassian models are shown, all with $\Omega_m^{obs} =
0.3$: $n=0.2$, $q=1$ (solid curve); $n=0.2$, $q=2$ (short-dashed
curve); and $n=0.2$, $q=3$ (long-dashed curve). 
Note that the solid
curve is equivalent to a quintessence model with $w_q = -0.8$. 
The dot-dashed curve is a quintessence model
with $w_q = -1 +0.5 z$.
The dotted curves show several familiar cosmological models
for comparison (from top to bottom): ($\Omega_m$, $\Omega_{\Lambda})=$
(0.3, 0.7); (0.3, 0); and (1, 0).}

\figcaption[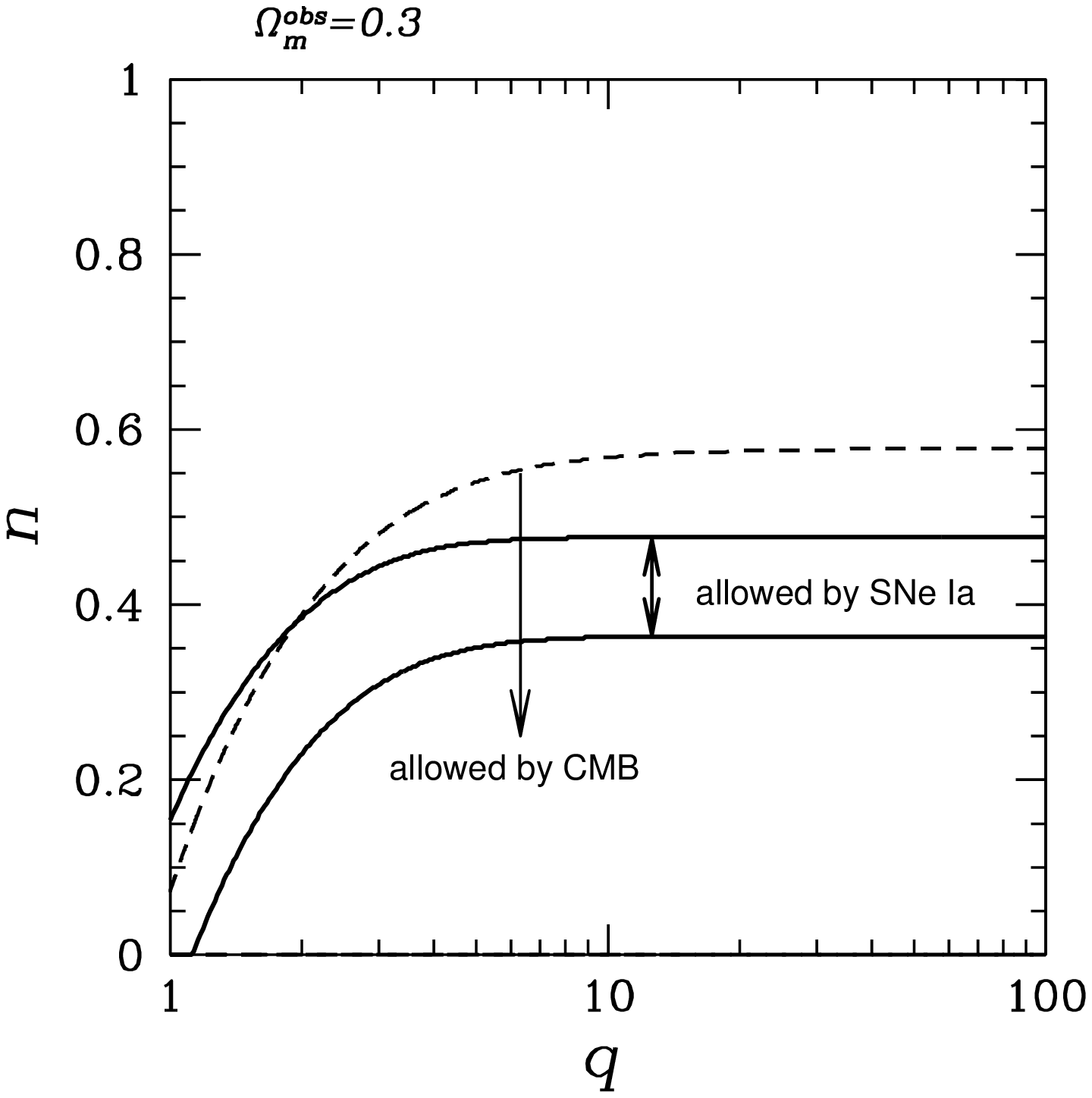]
{The parameter space of $(n,q)$ showing
 constraints from SNe Ia and CMB at fixed $\Omega_m=0.3$.  All models
below the dashed line are in agreement with the shift parameter 
${\cal R}$ as 
measured by WMAP.  All models between the solid lines are in agreement
with the Type Ia SN data.
}

\figcaption[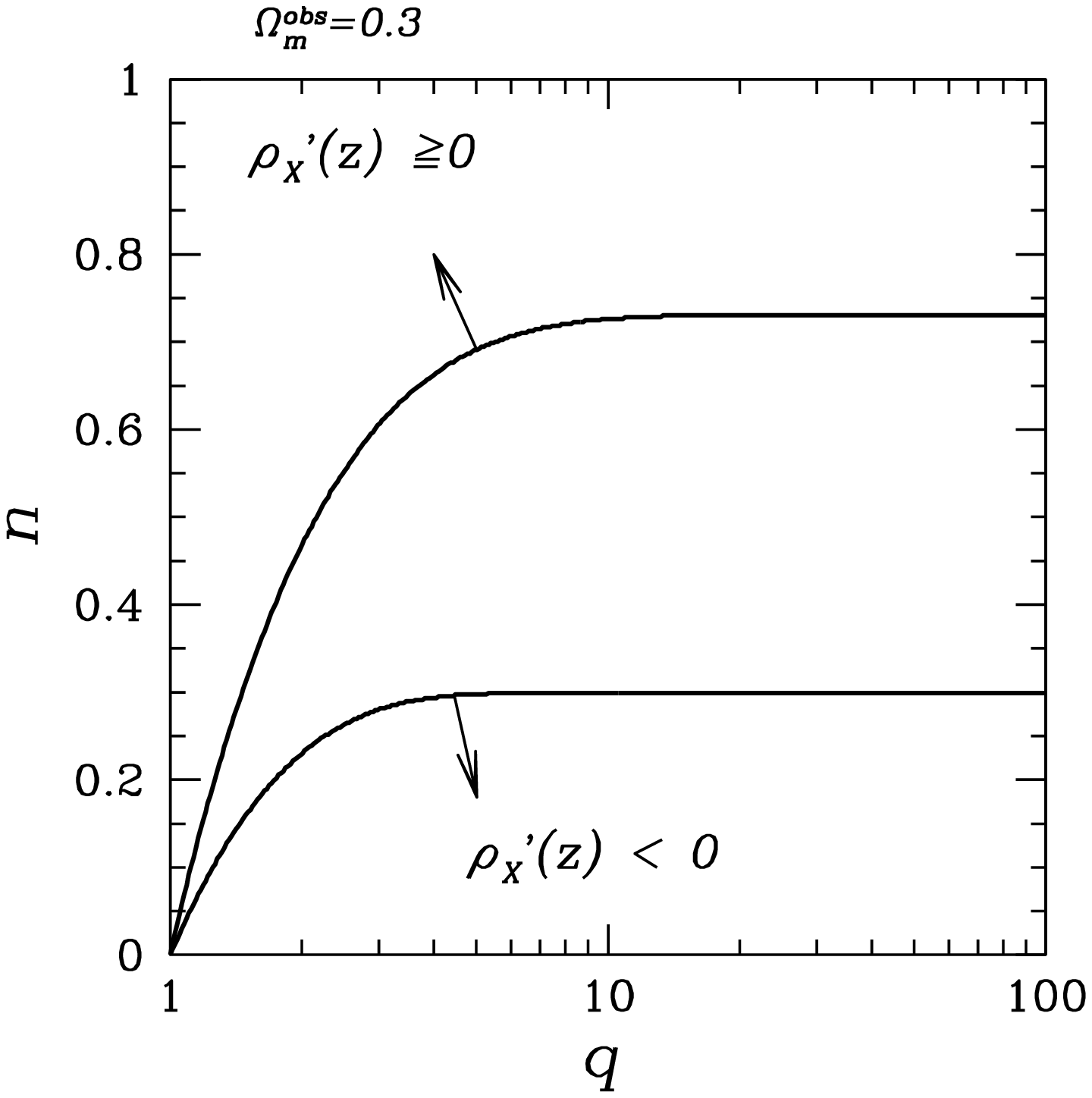]
{The parameter space
of (n,q) for MP Cardassian models [see Eq.(\ref{eq:FRW})]
with $\Omega_m^{obs}=0.3$. The arrows indicate the regions
in which $\rho_X'(z) \geq 0$ and $\rho_X'(z) < 0$ respectively.
The region
in between 
indicates models with dark energy densities that are {\it not}
monotonic functions of time.}

\figcaption[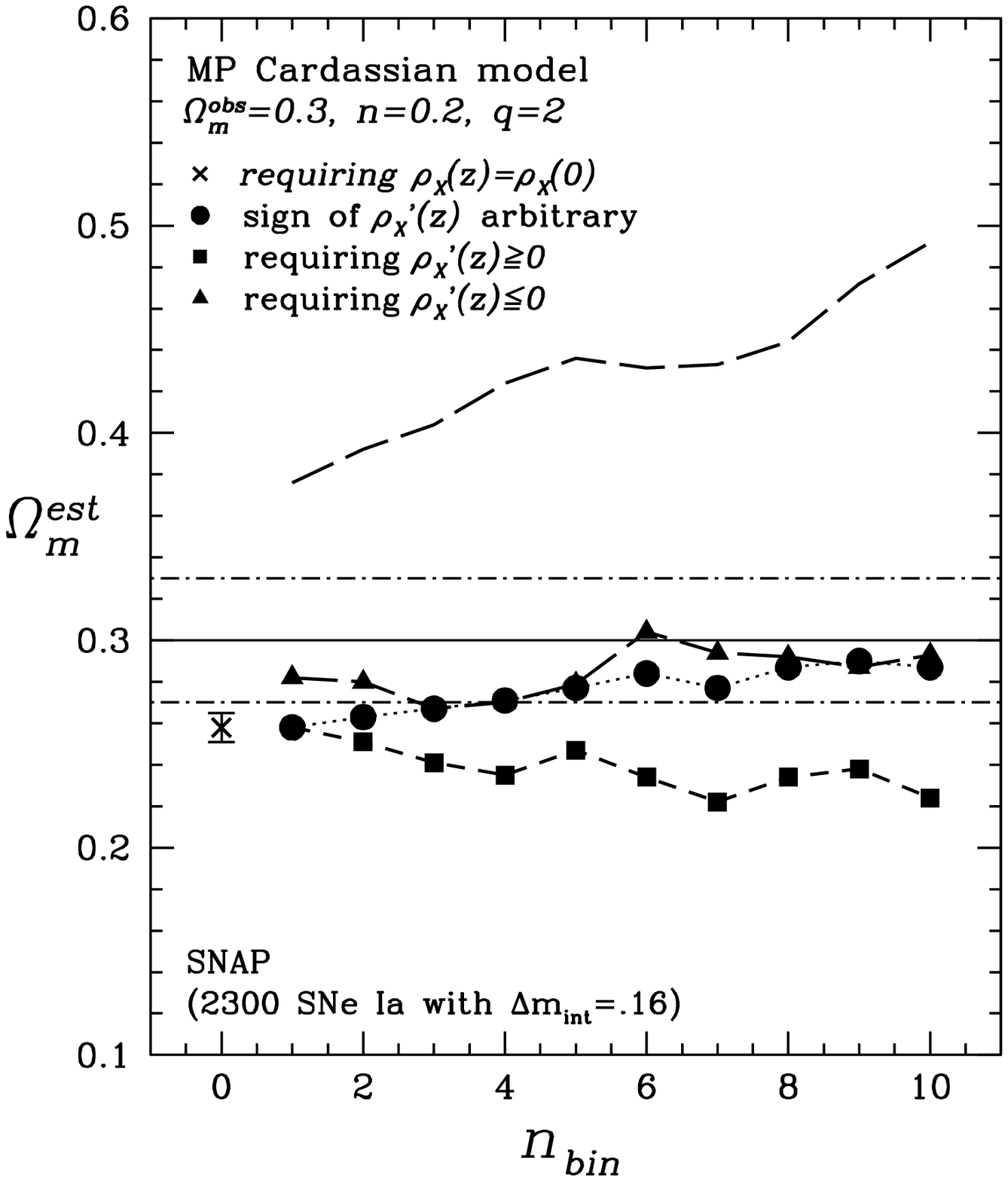]
{This figure shows that we can indeed determine
the sign of the dependence of the dark energy density 
$\rho_X'(z)$; i.e., we can determine if it is increasing,
decreasing, or constant in time.  The axes show
estimated $\Omega_m^{est}$ as function 
of $n_{bin}$ from the simulated data for SNAP
for (a) a MP Cardassian model with $n=0.2$ and $q=2$
so that $\rho_X'(z) < 0$
and (b) a quintessence model with $w_X(z)=-1+0.5z$ so that 
$\rho_X'(z) >0$.
The horizontal dot-dashed lines correspond
to a 10\% uncertainty on $\Omega_m^{obs} = 0.3 \pm 0.03$.
The different curves show the results obtained assuming
a variety of constraints 
on the time dependence $\rho_X'(z)$ as labeled.  
The long-dashed curve shows $\chi^2_{pdf}$ on an arbitrary scale.
By requiring
the results to lie within the dot-dashed lines, we recover
the sign of the time-dependence of $\rho_X(z)$.
}

\figcaption[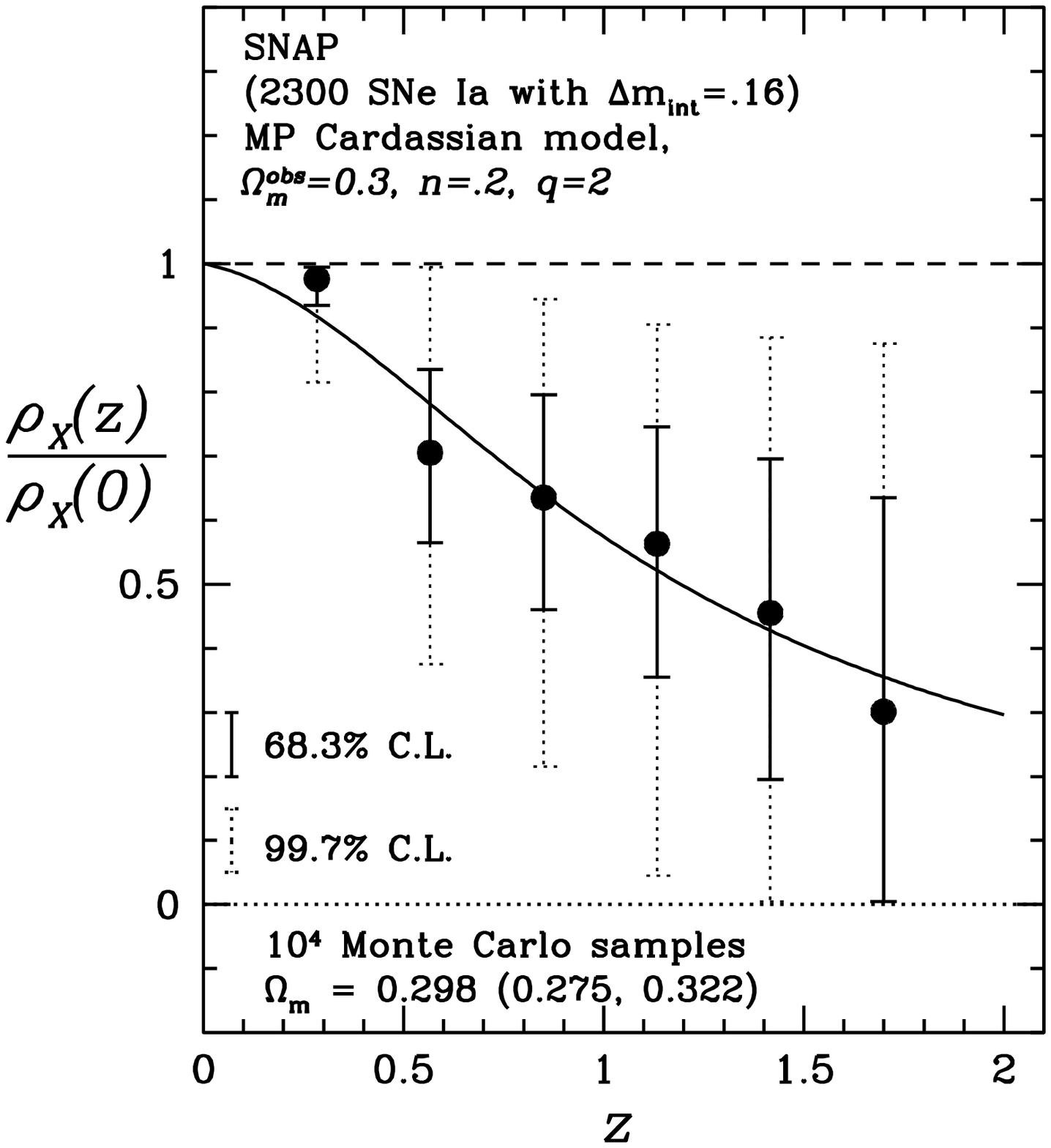]
{Estimated dimensionless dark energy density
$\rho_X(z)$ for simulated SN data from SNAP, assuming that we know 
$\Omega_m$ to $\sim$10\% accuracy
[and the correct general time
dependence of the dark energy density, see discussion
in Section 3.A]. 
The error bars of the estimated $\rho_X(z)$ have been computed from
$10^4$ Monte Carlo random samples derived from the simulated data.
The solid error bars and the dotted error bars 
indicate the 68.3\% and 99.73\% confidence level intervals respectively.
The reproduced estimates of 
$\Omega_m$ values are listed at the bottom of the plot
 with 68.3\% confidence level 
intervals.}

\clearpage
\setcounter{figure}{0}
\plotone{f1.eps}
\figcaption[f1.eps]
{Examples of MP Cardassian models [see
Eq.(\ref{eq:FRW})] that satisfy current observational
constraints from type Ia
supernovae (SNe Ia) data.
Three MP Cardassian models are shown, all with $\Omega_m^{obs} =
0.3$: $n=0.2$, $q=1$ (solid curve); $n=0.2$, $q=2$ (short-dashed
curve); and $n=0.2$, $q=3$ (long-dashed curve). 
Note that the solid
curve is equivalent to a quintessence model with $w_q = -0.8$. 
The dot-dashed curve is a quintessence model
with $w_q = -1 +0.5 z$.
The dotted curves show several familiar cosmological models
for comparison (from top to bottom): ($\Omega_m$, $\Omega_{\Lambda})=$
(0.3, 0.7); (0.3, 0); and (1, 0).}

\plotone{f2.eps}
\figcaption[f2.eps]
{The parameter space of $(n,q)$ showing
 constraints from SNe Ia and CMB at fixed $\Omega_m=0.3$.  All models
below the dashed line are in agreement with the shift parameter 
${\cal R}$ as 
measured by WMAP.  All models between the solid lines are in agreement
with the Type Ia SN data.
}

\plotone{f3.eps}
\figcaption[f3.eps]
{The parameter space
of (n,q) for MP Cardassian models [see Eq.(\ref{eq:FRW})]
with $\Omega_m^{obs}=0.3$. The arrows indicate the regions
in which $\rho_X'(z) \geq 0$ and $\rho_X'(z) < 0$ respectively.
The region
in between 
indicates models with dark energy densities that are {\it not}
monotonic functions of time.}

\plottwo{f4a.eps}{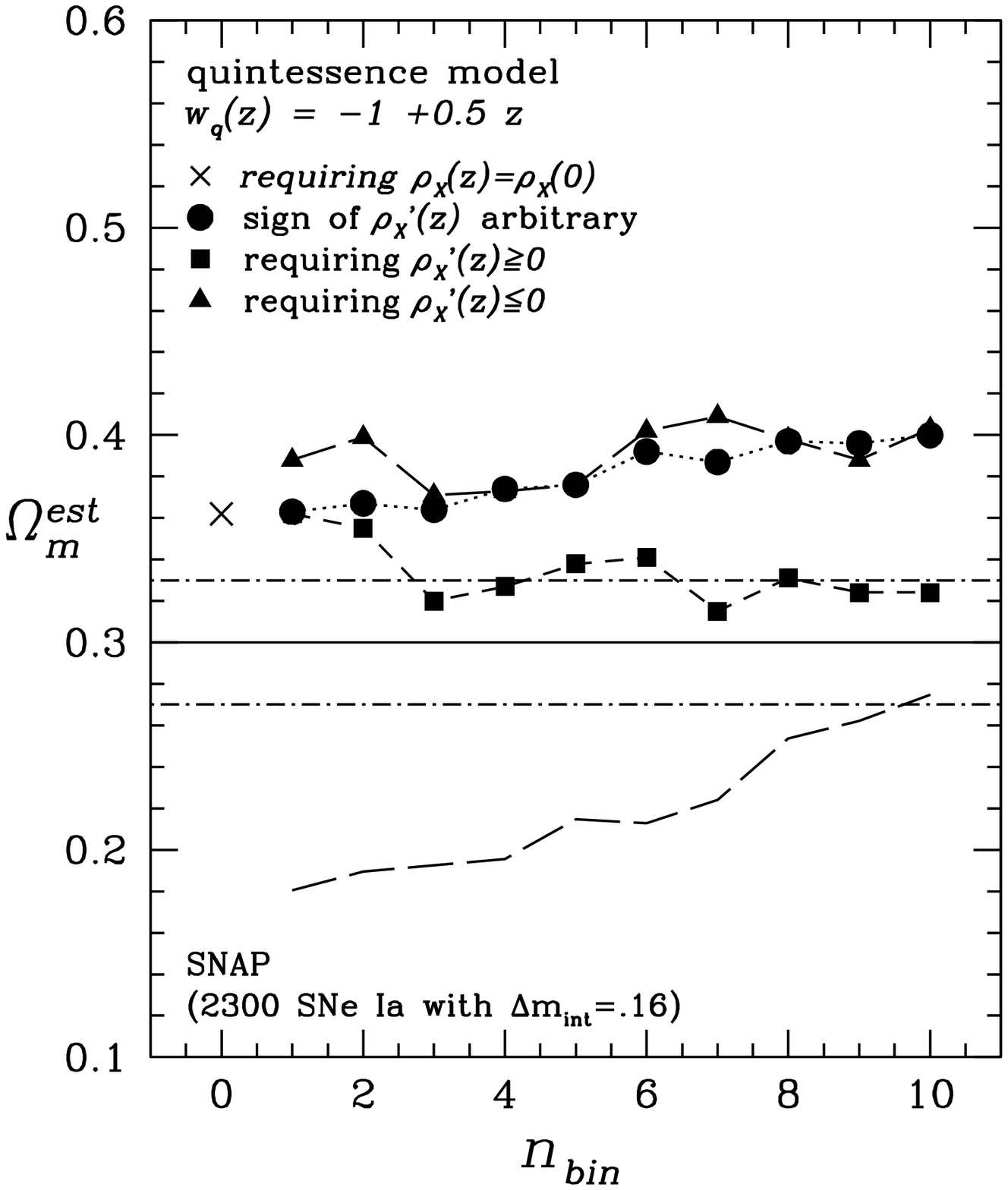}
\figcaption[f4a.eps]
{This figure shows that we can indeed determine
the sign of the dependence of the dark energy density 
$\rho_X'(z)$; i.e., we can determine if it is increasing,
decreasing, or constant in time.  The axes show
estimated $\Omega_m^{est}$ as function 
of $n_{bin}$ from the simulated data for SNAP
for (a) a MP Cardassian model with $n=0.2$ and $q=2$
so that $\rho_X'(z) < 0$
and (b) a quintessence model with $w_X(z)=-1+0.5z$ so that 
$\rho_X'(z) >0$.
The horizontal dot-dashed lines correspond
to a 10\% uncertainty on $\Omega_m^{obs} = 0.3 \pm 0.03$.
The different curves show the results obtained assuming
a variety of constraints 
on the time dependence $\rho_X'(z)$ as labeled.  
The long-dashed curve shows $\chi^2_{pdf}$ on an arbitrary scale.
By requiring
the results to lie within the dot-dashed lines, we recover
the sign of the time-dependence of $\rho_X(z)$.
}

\plotone{f5.eps}
\figcaption[f5.eps]
{Estimated dimensionless dark energy density
$\rho_X(z)$ for simulated SN data from SNAP, assuming that we know 
$\Omega_m$ to $\sim$10\% accuracy
[and the correct general time
dependence of the dark energy density, see discussion
in Section 3.A]. 
The error bars of the estimated $\rho_X(z)$ have been computed from
$10^4$ Monte Carlo random samples derived from the simulated data.
The solid error bars and the dotted error bars 
indicate the 68.3\% and 99.73\% confidence level intervals respectively.
The reproduced estimates of 
$\Omega_m$ values are listed at the bottom of the plot
 with 68.3\% confidence level 
intervals.}

\end{document}